\pdfoutput=1
\pdfoutput=1
\pdfoutput=1
\pdfoutput=1
\pdfoutput=1

\documentclass[fleqn]
{article}

\usepackage[T1]{fontenc}
\usepackage{amsfonts,amssymb,amsmath} % for math symbols.
\usepackage{bm}% bold math
\usepackage[all]{xy}
\usepackage{graphics,graphicx,epsfig}
\usepackage{color}
\usepackage{fancyhdr}
\fancypagestyle{plain}{%
 \fancyhead{} % ----на обычных страницах без колонтитулов

}
\usepackage[cp1251]{inputenc}
\usepackage [ english
] { babel }
\usepackage{makeidx}
\makeindex

\usepackage{xr-hyper}
\usepackage[unicode,pagebackref]{hyperref}
\hypersetup{pdftex}

\newcommand{\beq}{\begin{equation}}
\newcommand{\eeq}{\end{equation}}
\newcommand{\beqa}{\begin{eqnarray}}
\newcommand{\eeqa}{\end{eqnarray}}

 \title{
 The blind detection for palette image watermarking without changing the color
 }
 \author{V.N. Gorbachev, E.M. Kaynarova \footnote{E-mail:  helenkainarova@gmail.com}, I.K. Metelev, O.V. Pavlovskaya}
\date{\emph{\small {North-Western Institute of Printing}}\\
\emph{\small{St.-Petersburg State University of Technology and Design}} }

\makeindex
\begin{document}
\maketitle
\begin{abstract}
To hide a binary pattern in the palette image a steganographic scheme with blind detection is considered.
The embedding algorithm uses the Lehmer code by palette color permutations for which the cover image palette is generally required. The found transformation between the palette and RGB images allows to extract the hidden data without any cover work.
\end{abstract}

\section{Introduction}
Palette or indexed images are widespread in the Internet and various steganographic techniques have been developed to protect them. Graphic formats store the palette image creating a data set and a palette array
which can be used for hiding the information.
The modification of data set or palette as LSB-embedding often introduces noticeable artifacts. The possible solutions include sorting palette, parity and adaptive embedding \cite{1} and others \cite{1.1}.

Numerous techniques are based on the fact that reordering the palette colors and reindexing the data set can leave the image visibility unchanged. The GIF-Shuffle steganography program
% by Matthew Kwan
implements this solution for GIF images \cite{2}.

The GIF-Shuffle encodes message using  the  palette color permutations  and uses an algorithm that generates permutation according to a given integer. In the same manner SteganoGifPaletteOrder uses permutations of colors in the GIF palette \cite{3}.
Together with color images such technique, known as GIF-it-Up, allows to hide information in grayscale GIF images \cite{4}.

Encoding by permutation also known as Lehmer code \cite{LEHcode} assumes a choice of initial ordering. For instance, the GIF-Shuffle program sorts the palette RGB colors of the GIF using a ''natural'' order, i.e. every color is assigned to the value $6536R+256G+B$ and then sorted accordingly. It means that to retrieve the hidden information it needs this ''natural'' order or in other words it needs the cover work. From this point of view the steganographic system used the permutation code would have not blind detection. This is not a fatal restriction. Such ''natural'' order may be considered as a secrete key. The key can be distributed among the legitimate users via numerous classic cryptographic protocols and via a quantum protocol such as BB84 \cite{BB84}.

In this paper we propose a scheme with blind detection. The reason is that the cover image may be generated from the palette stego image if it will could be transformed into its RGB representation. That is possible thanks to the particular transformation between the palette and the RGB images. The main aim of our paper is to consider this map and to discuss the embedding of a binary pattern using the Lehmer code.

To improve the security level the GIF-Shuffle and the SteganoGifPaletteOrder techniques propose encryption. A possible encryption algorithm may be the ICE (Information Concealment Engine) with a 64 bit key \cite{5.1}. Indeed, to hide binary images a Vernam cipher may be suitable. When the key is implemented by a random binary matrix a one-time pad can be built. This is a perfect encryption if it is used correctly \cite{Shannon}. On practice one-time pad application is limited, it needs storing and distributing a large random key, that is very expansive. However its modern implementation based on the optical beam splitter of random structure may be efficient \cite{6}.

The paper is organized as follows: at first the Lehmer code is briefly discussed, then transformations of palette images are considered and finally a steganographic scheme with blind detection is described.

\section{Lehmer code}

The Lehmer code enumerates permutations by an integer represented in the factorial number system . %\cite{LEHcode}.
It allows encoding the binary message by a permutation and it needs an initial permutation called identity.
\\
\\
 Positive integer $m$ can be represented with the help of the factorial number system as
 \begin{eqnarray}
\label{100}
 m=\sum_{k=1}^{n}a_{k}(k-1)!,
\end{eqnarray}
where factorial digits
$a_{k}= \lfloor m/(k-1)!\rfloor \mod(k)$ take their value from the set $0,1,2,\dots k-1$, and the floor function $\lfloor x\rfloor $ computes the largest integer less or equal to $x$. The representation (\ref{100}) can be considered as a factorial $n$-digital register, that stores any integer from the range $[0, n!-1]$. For example, six factorial digits of 251 from $a_{6}$ to $a_{1}$ read
\begin{eqnarray}
\label{101}
 251= (2, 0, 2,0,1,0)_{!}=2\cdot 5!+0\cdot 4!+2\cdot 3!+0\cdot 2!+1\cdot 1!+0\cdot 0!.
\end{eqnarray}
The Lehmer code is one of the ways of encoding permutations using the inversion.
Items $x_{k}$ and $x_{j}$ of a permutation are called in the inversion if $x_{k}> x_{j}$ for $k<j$. Let $t_{k}$ be the number of all inversions of $x_{k}$, then vector
$
t=
(t_{n}, t_{2},\dots t_{1})
$
 can uniquely describe any permutation of $n$ entries. As there are $n-k$ items to the right from $x_{k}$, then $x_{k}$ creates no more than $n-k$ inversions and $t_{k}$ can take its value $0,1,2\dots n-k$, where 0 denotes no inversion and $\max t_{k}=n-k$.
 \\
 For example, assume the permutation of $n=6$ entries be $P(012345)=204153$, where $012345$ is the identity permutation. Then $\max t=(5,4,3,2,1,0)$, where 0 indicates the fact that the item 5 is the last and there are no items on its right. For this case
\begin{eqnarray}
\label{102}
 \begin{array}{rcccccc}
 P(012345) \ =& 2 & 0 & 4 & 1 & 5 & 3 \\
 t= & (2, & 0, & 2, & 0, & 1, & 0 ),
 \end{array}
\end{eqnarray}
where the first item 2 from $t$ indicates that the first item of permutation 2 creates inversions with 0 and 1.

The considered examples (\ref{101}) and (\ref{102}) illustrate the one-to-one correspondence between the permutation presented by its inversions and the integer $m$ presented by its factorial digits
\begin{eqnarray*}
251= (2, 0, 2,0,1,0)_{!}\leftrightarrow 204153.
\end{eqnarray*}
Another example is the permutation of $\max t=(n-1,\dots,2,1,0)$ that refers to full inversion of its entries $P_{inv}(012\dots n-1)=n-1\dots210$
\begin{eqnarray}
\label{3}
n!-1=(n-1,\dots,2,1,0)_{!}\leftrightarrow n-1\dots210.
\end{eqnarray}

\begin{figure}
 \includegraphics[height=14cm,width=16cm]{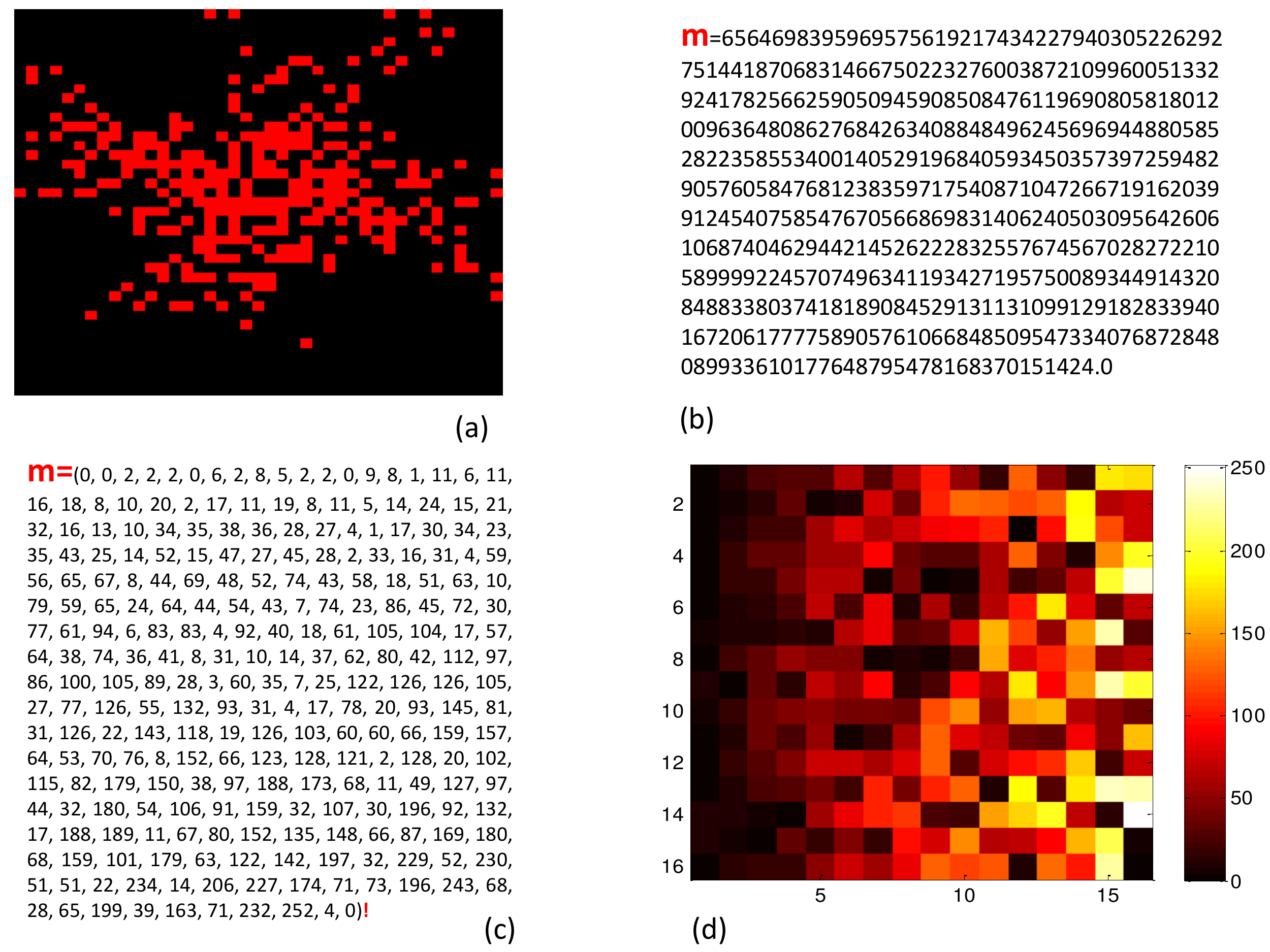}\\
 \caption{Encoding of binary image by the factorial digits of the integer:
 binary image of $41\times 41$ pixels (a); the 503 digital integer corresponding to the binary image (b); the 256 factorial digits of the integer $m$ (c); the $16\times16$ matrix representing the factorial digits (d).
 }
 \label{f0}
\hyperlink{f0}{W}
\end{figure}

The Lehmer code allows to encode any digital images using permutations. For instance, considering a binary image, this is a set of binary digits of an integer $m$ that can be represented in the factorial number system.
\hypertarget{f0}
Figure (\ref{f0}) illustrates first steps. A $41\times 41$ binary image of two colors (red and black) is transformed into an integer $m$ that consists of 503 decimal and 256 factorial digits. Factorial digits are arranged into the $16\times16$ matrix shown in the Fig. (\ref{f0}) (d).
 \\
Then to generate the desired permutation for the given $m$ an identity permutation has to be selected.
This fact is important for the palette color embedding techniques
based on the color permutation. For this case the identity permutation is given by the cover palette and it is required for detection of hidden data. As result the detection algorithm will use the cover image and will not be blind. However the cover palette can be achieved from a stego image thanks to particular features of the palette color transformation.

\section{Transformation between palette and color images}

A palette image can be converted into another one so that both images will have the same RGB appearance that can be converted reversibly only to one of them.
\\
\\
From the mathematical point of view a palette or indexed image can be represented by a tensor product $C_{pal}=I\otimes P$, where $I$ is a data set or an index array, and $P$ is a palette that maps the color in RGB model. When index $i\in I$ refers to the palette it takes the color and then there is a transformation $C_{pal} \to C_{rgb}$, where $C_{rgb}$ is an RGB representation of the palette image or its appearance we see on the display.
\\
A common transformation between color and palette images is irreversible.
The usual representation has 24 bit for color pixel and 8 bit for palette pixel. A large number of color quantization algorithms allows us to convert a 24-bit color image into a palette image of 8 bit color - $C_{pal}(8)$. Here the entropy is not preserved and this transformation is lossy, however the achieved palette image may look nice.
\\
 Let the image $C_{pal}(8)$ have its RGB appearance of 8 bit color $C_{rgb}(8)$. Using the same color quantization algorithms we convert this RGB image back into the palette one. The algorithms are assumed to be deterministic then clear that output is $C_{ind}(8)$. It means that there is the one-to-one correspondence between $C_{rgb}(8)$ and $C_{ind}(8)$:
 \begin{eqnarray}
\label{1}
 %A: \
 C_{rgb}\rightleftarrows C_{ind}.
\end{eqnarray}
Indeed, this reversible transformation can not couple all pairs of images and there is a set of palette images for which (\ref{1}) is not reversible.

Palette images can be transformed from one to another using different approaches.
A simultaneous changing of the index and the palette colors may result in a transformation
$ C_{pal}\to S_{pal}$, where both palette images have the same RGB appearance $C_{rgb}$.
 Let $C_{pal}$ and $C_{rgb}$ be coupled by (\ref{1}). Then clear that the quantization algorithms convert $C_{rgb}$ into $C_{pal}$ instead of $S_{pal}$. This observation is described by the next mapping
\begin{eqnarray}
\label{2}
 C_{rgb}\rightleftarrows C_{pal}\to S_{pal}\to C_{rgb}.
 \end{eqnarray}
Figure (\ref{f1}) illustrates \hypertarget{f1} the map given by (\ref{2}). Here the RGB color image $C_{rgb}$
and the palette image $C_{pal}$ are converted from one to another by Matlab functions
 \texttt{rgb2ind} and \texttt{ind2rgb} in accordance with
 Eq (\ref{1}). Two palette images are given by its index arrays, grayscale patterns and palettes.
 The image $S_{pal}$ is a negative of $C_{pal}$. The negative is achieved by replacing the indexes and the palette colors or the palette rows with $i\to i'=255-i$.
Being negative patterns both images $C_{pal}$ and $S_{pal}$ have the same RGB representation given by $C_{rgb}$.

\section{Steganographic scheme with blind detection}

\begin{figure}
 \includegraphics[height=14cm,width=16cm]{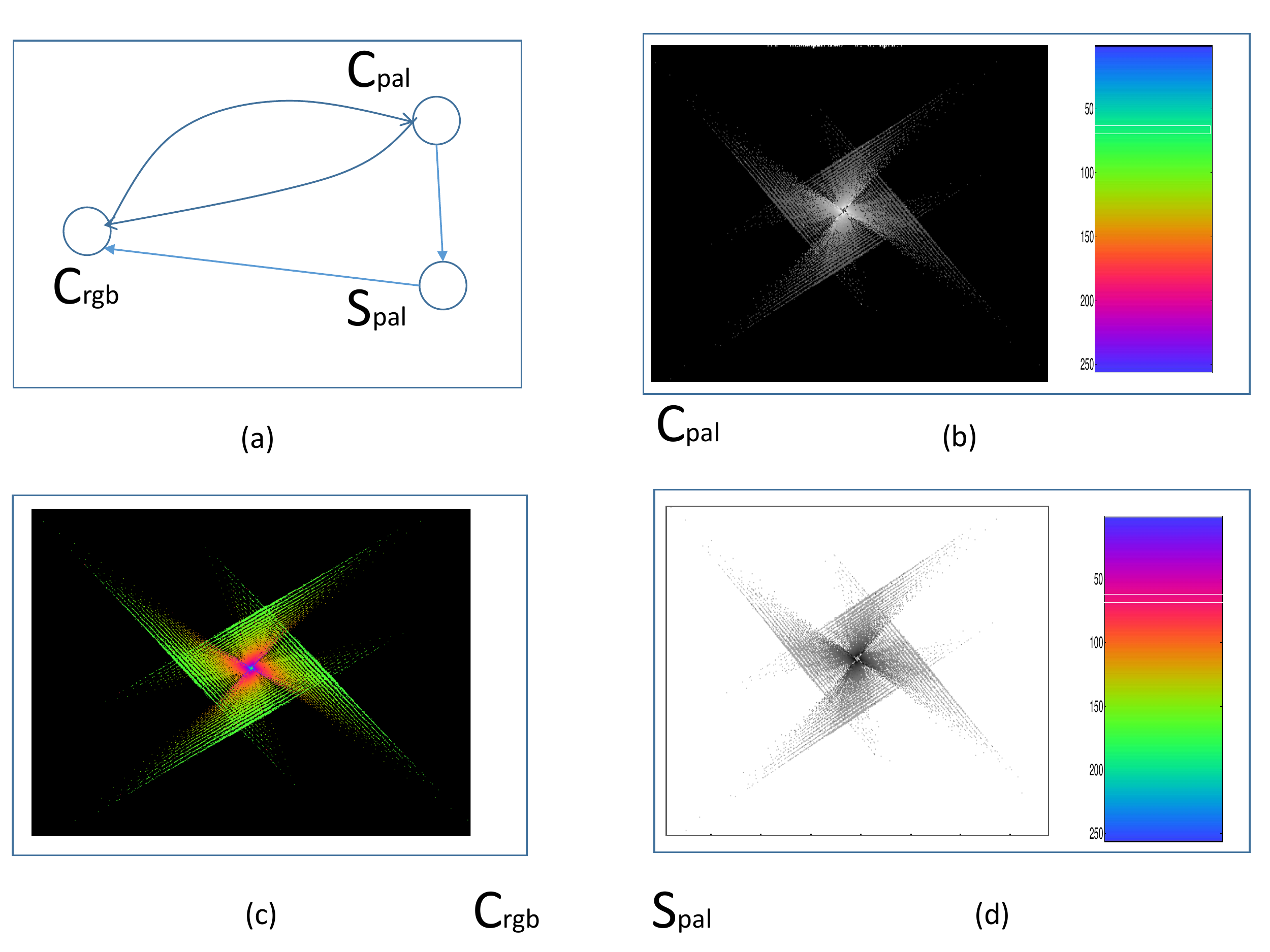}\\
 \caption{Transformations between the color and palette images (a); palette image $C_{pal}$ and its negative $S_{pal}$ presented by its data sets, grayscale images, and palettes (b) and (d); these palette images have the same RGB appearance $C_{rgb}$ (c).
 }
 \label{f1}
\hyperlink{f1}{W}
\end{figure}

 Two palette images that have the same RGB appearance may be considered as a pair of cover and stego works of a steganographic scheme where the hidden message can be extracted from the stego image.
 \\
 \\
Two palette images shown in the Fig (\ref{f1}) have the same look and one of them - $S_{pal}$ is achieved from $C_{pal}$ by inversion of its colors and indexes. Using this approach a message was embedded and it can be established easily. It follows from the Lehmer encoding that the permutation that inverts all $n$ colors has integer $m=n!-1$ (\ref{3}). This integer consists of not less than $ N=\log_{2}(n!)$ binary digits and is a message $M$ embedded into the palette.
Without loss of generality we will consider $M$ as a binary image. However this image will consist of 1 only and its Shannon entropy $H(M)=0$. For a palette of $n=256$ colors $N\geq 1684$ and being accurate up to 3 bits we find that $M$ is a $41\times 41$ matrix of 1 that may look as a black square. It means that embedding the message as a "black square'' is a complete inversion of the palette colors.

\begin{figure}
 \includegraphics[height=11cm,width=16cm]{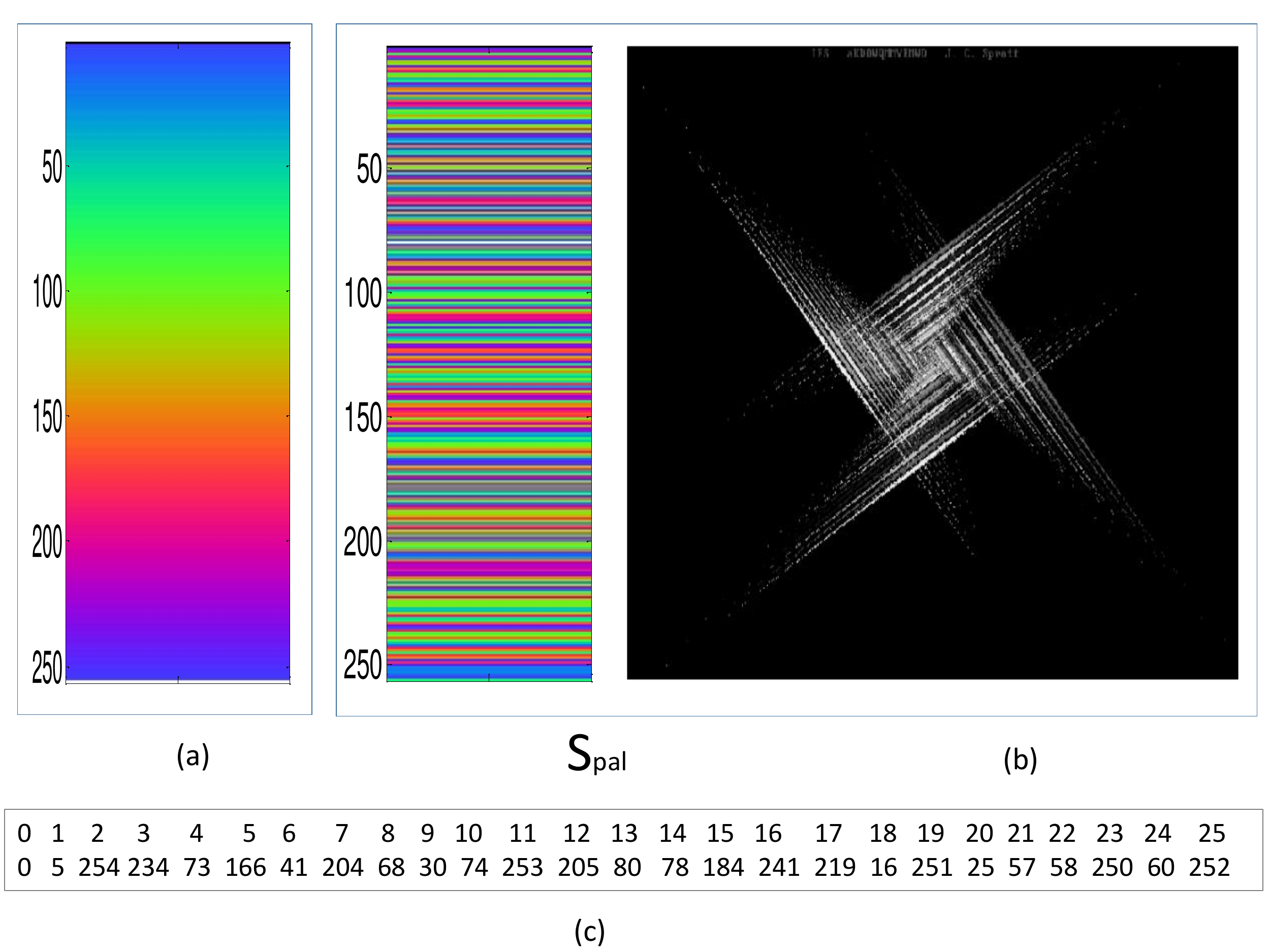}\\
 \caption{Embedding of binary image by the Lehmer code. Binary image and cover work shown in the Fig. (\ref{f1}) (b) and in the Fig. (\ref{f0}) (a). The cover image palette (a); palette stego image achieved by permutation of the colors and indexes (b); a piece of the permutation (c).
 }
 \label{f2}
\hyperlink{f2}{W}
\end{figure}

 \hypertarget{f2}
 Fig. (\ref{f2}) illustrates embedding a binary image by the Lehmer code. The
 $41\times 41$ binary image presented in the Fig. (\ref{f0}) (a) is embedded into the cover work $C_{pal}$ in the Fig. (\ref{f1}) (b). The stego image $S_{pal}$ is shown in the Fig. (\ref{f2}) (b), and is achieved by the found permutation of the palette colors and indexes of $C_{pal}$. A piece of the permutation is given at the button, in the Fig. (\ref{f1}) (c).

After the message is embedded the stego image $S_{pal}$ can be storied in such graphic formats as GIF and PNG for further use without any lossy transformations. To extract the message from $S_{pal}$ it should be downloaded from GIF or PNG and in accordance to (\ref{2}) the solution is obtained after executing the following steps:
\begin{itemize}
 \item palette is extracted from $S_{pal}$ and cover image $C_{pal}$ is generated from $C_{rgb}$ that is the color representation of both the stego and the cover images;
 \item palette is extracted from $C_{pal}$ that is the identity permutation of colors;
 \item the hidden message is extracted from the the palettes by Lehmer encoding.
\end{itemize}

 Note that the presented scheme is not secrete from the Kerckhoff's principal point of view because it has no secrete key. A possible solution is to generate a random binary matrix of the size equal to the binary message and combine them, for example, by modular addition. When the matrix is random and is never reused the obtained system (known also as a one-time pad) is unbreakable.

\section{Acknowledgments}
We are grateful to Volker Jansen for attention to our work. One of the authors (E.M.K.) was supported by DAAD (German Academic Exchange), under grant No 11.9170.2014.

%\newpage
%\input{reFs}
%\end{document}
%%%%%%%%%%%%%%%%%%%%%%%%%%%%
%%%%%%%%%%%%%%%%%%
%%%%%%%%%%%%%%%%%%%%%%%%%%%%%%%
%%%%%%%%%%%%%%%%%%%

\begin{thebibliography}{99}
\bibitem{1} J. Fridrich, R. Dui. Secure steganographic methods for palette images. In 3rd Information Hidding Workshop, Lecture note on computer science (1768), New York, Heidelberg, 1999, Spriger-Verlag.
    \\
    R. Machado. EZ Stego, Stego Online, Stego, http://www.stego.com, 2003.
    \par
   \bibitem{1.1}
    C-C Chang,  C-C Lin, Y-H Chen. Hiding data in colour palette images with hybrid strategies,  The Imaging Science Journal, Volume 57, Issue 1 (01 February 2009), pp. 19-29.
\par
 \bibitem{2} http://www.darkside.com.au/gifshuffle/
    \par
     \bibitem{3} % SteganoGifPaletteOrder
   http://users.skynet.be/glu/sgpo.htm  %(2002)
    \par
    \bibitem{4}
     http//www.theargon.com/achilles/steganography/gif-it-up/ % не работает
    \\
    Shiguo Lian, Yan Zhang. Handbook of Research on Secure Multimedia Distribution. IGI Global, 2011.
        \par
\bibitem{LEHcode}
D. H. Lehmer, Teaching combinatorial tricks to a computer, Proc. Sympos. Appl. Math. Combinatorial Analysis, Vol. 10, Amer. Math. Soc., Providence, R. I., 179-193, (1960).
\par
 \bibitem{BB84}C.H. Bennett,  G. Brassard. Quantum public key distribution system. IBM Tech. Discl. Bull. 28, 3153- 3163
(1985).
\par
\bibitem{5.1} http://www.darkside.com.au/ice/index.html
\par
  \bibitem{Shannon} C. Shannon.  Communication theory of secrecy systems. Bell System Technical Journal 28 (4), 656715 (1949).
      \par
\bibitem{6}
R.  Horstmeyer, B. Judkewitz,  I. Vellekoop,  S.  Assawaworrarit  and C.  Yang.
Physical key-protected one-time pad. arXiv: 1305.3886v2 \ [physics.optics] 26 Nov 2013, (2013).
\par

\end{thebibliography}
\end{document}